\documentclass[12pt]{article}
\usepackage{graphicx}
\usepackage{lscape}
\usepackage{citesort}

\begin{document}

\def\ds{\displaystyle}
\def\beq{\begin{equation}}
\def\eeq{\end{equation}}
\def\bea{\begin{eqnarray}}
\def\eea{\end{eqnarray}}
\def\beeq{\begin{eqnarray}}
\def\eeeq{\end{eqnarray}}
\def\ve{\vert}
\def\vel{\left|}
\def\ver{\right|}
\def\nnb{\nonumber}
\def\ga{\left(}
\def\dr{\right)}
\def\aga{\left\{}
\def\adr{\right\}}
\def\lla{\left<}
\def\rra{\right>}
\def\rar{\rightarrow}
\def\nnb{\nonumber}
\def\la{\langle}
\def\ra{\rangle}
\def\ba{\begin{array}}
\def\ea{\end{array}}
\def\tr{\mbox{Tr}}
\def\ssp{{\Sigma^{*+}}}
\def\sso{{\Sigma^{*0}}}
\def\ssm{{\Sigma^{*-}}}
\def\xis0{{\Xi^{*0}}}
\def\xism{{\Xi^{*-}}}
\def\qs{\la \bar s s \ra}
\def\qu{\la \bar u u \ra}
\def\qd{\la \bar d d \ra}
\def\qq{\la \bar q q \ra}
\def\gGgG{\la g^2 G^2 \ra}
\def\q{\gamma_5 \not\!q}
\def\x{\gamma_5 \not\!x}
\def\g5{\gamma_5}
\def\sb{S_Q^{cf}}
\def\sd{S_d^{be}}
\def\su{S_u^{ad}}
\def\ss{S_s^{??}}
\def\sbp{{S}_Q^{'cf}}
\def\sdp{{S}_d^{'be}}
\def\sup{{S}_u^{'ad}}
\def\ssp{{S}_s^{'??}}
\def\sig{\sigma_{\mu \nu} \gamma_5 p^\mu q^\nu}
\def\fo{f_0(\frac{s_0}{M^2})}
\def\ffi{f_1(\frac{s_0}{M^2})}
\def\fii{f_2(\frac{s_0}{M^2})}
\def\O{{\cal O}}
\def\sl{{\Sigma^0 \Lambda}}
\def\es{\!\!\! &=& \!\!\!}
\def\ap{\!\!\! &\approx& \!\!\!}
\def\ar{&+& \!\!\!}
\def\ek{&-& \!\!\!}
\def\kek{\!\!\!&-& \!\!\!}
\def\cp{&\times& \!\!\!}
\def\se{\!\!\! &\simeq& \!\!\!}
\def\eqv{&\equiv& \!\!\!}
\def\kpm{&\pm& \!\!\!}
\def\kmp{&\mp& \!\!\!}


\def\simlt{\stackrel{<}{{}_\sim}}
\def\simgt{\stackrel{>}{{}_\sim}}


\title{
         {\Large
                 {\bf
Unparticle physics effects in $\Lambda_b \rar \Lambda+missing~ energy$
processes 
                 }
         }
      }

\author{\vspace{1cm}\\
{\small T. M. Aliev \thanks
{e-mail: taliev@metu.edu.tr}~\footnote{permanent address:Institute
of Physics,Baku,Azerbaijan}\,\,,
M. Savc{\i} \thanks
{e-mail: savci@metu.edu.tr}} \\
{\small Physics Department, Middle East Technical University,
06531 Ankara, Turkey} }

\date{}

\begin{titlepage}
\maketitle
\thispagestyle{empty}

\begin{abstract}
We study unparticle physics effects in $\Lambda_b \rar \Lambda+missing~
energy$ decay with polarized $\Lambda_b$ and $\Lambda$ baryons. The
sensitivity of the branching ratio of this decay and polarizations 
of $\Lambda_b$ and $\Lambda$ baryons on the scale dimension $d_{\cal U}$ and
effective cut--off parameter $\Lambda_{\cal U}$ are discussed. 
\end{abstract}

\end{titlepage}

\section{Introduction}

Flavor changing neutral current (FCNC) decays induced by the $b\rar s$
transition are promising decays for checking predictions
of the Standard model (SM) at quantum loop level, since they are forbidden at tree
level in the SM. These transitions are also very suitable in looking for new
physics beyond the SM.

In the SM the $b \rar s \nu \bar{\nu}$ decay receives special attention due
to the theoretical advantage that uncertainties in this decay are much
smaller compared to other FCNC decays due to the absence of photonic penguin
diagrams and hadronic long distance effects. However, in spite of these
theoretical advantages experimental measurement of this inclusive channel
seems to be very difficult, because it requires a construction of all
$X_s$. Therefore, experimentalists focus only on exclusive
channels like $B \rar K(K^\ast) \bar{\nu} \nu$. This channel studied
extensively on theoretical grounds in many works (see for example
\cite{R8701,R8702,R8703,R8704}). Another class of decays, which is described
by the $b \rar s \bar{\nu} \nu$ transition at inclusive level, is the
baryonic $\Lambda_b \rar \Lambda \bar{\nu} \nu$ decay.

It should be noted that it is impossible to analyze the helicity structure
of the effective Hamiltonian in $B$ meson decays governed by $b \rar s$
transition, since the information about chiralities of the quarks is lost in
the hadronization process. In contrary to the mesonic decays, baryonic decays
could access the helicity structure of the effective Hamiltonian for the
$b\rar s$ transition \cite{R8705}. Therefore the heavy baryonic decays can
be very rich for studying the polarization effects.

Radiative and semileptonic decays of $\Lambda_b$ baryon, such as $\Lambda_b
\rar \Lambda \gamma$, $\Lambda_b \rar \Lambda_c \ell \bar{\nu}_\ell$,
$\Lambda_b \rar \Lambda \ell^+ \ell^-$ and $\Lambda_b \rar \Lambda \bar{\nu}
\nu$, are comprehensively studied in the framework of SM in many works
\cite{R8705,R8706,R8707,R8708,R8709,R8710,R8711,R8712}. 
Present status of the experimental investigations of heavy baryons is
discussed in \cite{R8713}.

As has already been noted, FCNC transitions are very sensitive to the
existence of new physics beyond the SM. One such model is the so--called
unparticles is proposed by H. Georgi \cite{R8714}. It is assumed in this model
that, at very high energies the theory contains SM fields and the fields
with a non--trivial infrared fixed point so that in the infrared limit it will
be an asymptotic conformal theory and be scale--invariant, which are called
Banks--Zaks (BZ) fields \cite{R8715}. In unparticle physics model these two
sectors interacted by exchange of particle with a large mass scale $\mu$, below 
this scale where non--renormalizable couplings between SM and BZ fields will
be induced and renormalizable couplings between the BZ fields are then
produced by dimensional transmutation, and the scale--invariant unparticle
fields emerge below a scale $\Lambda_{\cal U}$. In the effective theory, below
$\Lambda_{\cal U}$, BZ operators match onto the unparticle operators, 
and non--renormalizable interactions between SM and unparticle operators
can be obtained. An important result in this theory is that unparticle stuff
with scale dimension $d_{\cal U}$ looks like a non--integer number $d_{\cal U}$ 
of invisible 
particles \cite{R8714}, where production might be detectable in missing
energy and momentum distributions. Various phenomenological aspects of the
unparticle physics have recently been extensively discussed in literature
\cite{R8716}--\cite{R8756}.

In the present work we study the $\Lambda_b \rar \Lambda+missing~ energy$
decay in unparticle physics. The paper is organized as follows: In section
2, we give necessary theoretical framework to describe the differential
decay width of $\Lambda_b \rar \Lambda+missing~energy$ in the SM and in
unparticle physics. Section 3 is devoted to numerical analysis and
conclusions.

\section{Theoretical framework}

In the SM $\Lambda_b \rar \Lambda+missing~energy$ channel is described by
the $\Lambda_b \rar \Lambda \bar{\nu} \nu$ decay. As has already been noted,
unparticles can also contribute to this decay. Therefore, a comparison of
the signature of the signature of the decay modes $\Lambda_b \rar \Lambda
\bar{\nu} \nu$ and $\Lambda_b \rar \Lambda U$ is required.

In the SM, the $\Lambda_b \rar \Lambda \bar{\nu} \nu$ decay is described at
quark level by the $b \rar s \bar{\nu} \nu$ transition and receives
contributions from $Z$--penguin and box diagrams, where main contributions
come from intermediate top quarks. The effective Hamiltonian responsible for
$b \rar s \bar{\nu} \nu$ transition is described by only one Wilson
coefficient $C_{10}$ and its explicit form is
\bea
\label{e8701}
{\cal H} = {G_F \over\sqrt{2}} {\alpha \over 2 \pi} V_{tb} V_{ts}^\ast
C_{10}
\bar{s} \gamma_\mu (1-\gamma_5) b \, \bar{\nu} \gamma^\mu (1-\gamma_5) \nu~,
\eea
where $G_F$ and $\alpha$ are the Fermi constant and structure constants,
respectively, $V_{ij}$ are the elements of the Cabibbo--Cobayashi--Maskawa
matrix (CKM). The Wilson coefficient $C_{10}$ in Eq. (\ref{e8701}),
including ${\cal O}(\alpha_s)$ corrections, has the following form:
\bea
\label{e8702}
C_{10} = {X(x_t) \over \sin^2\theta}~,
\eea
where 
\bea
\label{e8703}
X(x_t)  = X_0(x_t) + {\alpha_s \over 4 \pi} X_1(x_t)~.
\eea
$X_0(x_t)$ in Eq. (\ref{e8703}) is the usual Inami--Lim function
\cite{R8757} given by:
\bea
\label{e8704}
X_0(x_t) = {x_t \over 8} \Bigg[ {x_t+2 \over x_t-1} + {3 x_t-6 \over
(x_t-1)^2} \ln{x_t} \Bigg]~,
\eea
and
\bea
\label{e8705}
X_1(x_t) \es {4 x_t^3 - 5 x_t^2 - 23 x_t \over 3(x_t-1)^2} -
{x_t^4 + x_t^3 - 11 x_t^2 + x_t \over (x_t-1)^3} \ln{x_t} \nnb \\ 
\ar {x_t^4 - x_t^3 - 4 x_t^2 - 8 x_t \over 2 (x_t-1)^3} \ln^2{x_t}
+ {x_t^3 - 4 x_t \over (x_t-1)^2} Li_2 (1-x_t) +
8 x_t {\partial X_0(x_t) \over \partial x_t } \ln{x_\mu}~.
\eea
Here,
\bea
Li_2 (1-x_t) = \int_1^{x_t} dt {\ln t \over 1-t}~, \nnb
\eea
is the spence function, $x_t = m_t^2/m_W^2$, $x_\mu = \mu^2/m_W^2$ and
$mu$ describes the scale dependence when leading QCD corrections are taken
into account. The function $X_1(x_t)$ is calculated in \cite{R8758}.

Similarly, at quark level in unparticle physics, $b \rar s + missing~energy$
is described by the $b \rar s U$ transition, where we shall
consider two types of unparticle operators:
\begin{itemize}
\item Scalar unparticle operators,
\item Vector unparticle operators.
\end{itemize}

For the scalar and vector operators $b \rar s U$ transition is described by the
following matrix elements
\bea
\label{e8706}
&&{1 \over \Lambda_{d_{\cal U}}^{\cal U} } \Big[ \bar{s} \gamma^\mu (C_S + C_P \gamma_5 ) b
\Big] \partial_\mu {\cal O}_u~,\\
\label{e8707}
&& {1 \over \Lambda_{d_{\cal U}}^{d_{\cal U}-1} } \Big[ \bar{s} \gamma_\mu (C_V + C_A \gamma_5
) b \Big] {\cal O}_u^\mu~, 
\eea
where $C_i$ are the dimensionless effective couplings.

Before performing analytical calculations, let us present the forms of the
propagators for the scalar and vector unparticle physics \cite{R8716,R8717}
\bea
\label{e8708}
{\cal D} (q^2) \es \int d^4 x e^{iqx} \lla 0 \vel \mbox{\rm T} ({\cal O}_u(x)
{\cal O}_u(0) \ver 0 \rra~, \\
\es {A_{d_{\cal U}} \over 2 \sin (d_{\cal U} \pi) } (-q^2)^{d_{\cal U}-2}~,\nnb \\ \nnb \\
\label{e8709}
{\cal D}_{\mu\nu} \es {A_{d_{\cal U}} \over 2 \sin (d_{\cal U} \pi) } (-q^2)^{d_{\cal U}-2}
\Bigg( - g_{\mu\nu} + {q_\mu q_\nu \over q^2}\Bigg)~,
\eea
where
\bea
\label{e8710} 
A_{d_{\cal U}} = {16 \pi^{5/2} \over (2\pi)^{2 d_{\cal U}} } {\Gamma (d_{\cal U}+1/2) \over
\Gamma (d_{\cal U}-1)\Gamma (2d_{\cal U}) }~.
\eea
It is found in \cite{R8714} that, using scale invariance of the unparticle
physics, the phase for an unparticle operator with the scale dimension $d_{\cal U}$
and momentum $q$ is the same as the phase space for $d_{\cal U}$ invisible massless
particles
\bea
\label{e8711}
d\Phi_u(q) = A_{d_{\cal U}} \Theta(q^0) \Theta(q^2) (q^2)^{d_{\cal U}-2} {d^4q \over
(2\pi)^2}~.
\eea
Having the explicit forms of the effective Hamiltonian at hand for the $b
\rar \bar{\nu} \nu$ transition and effective interaction for the $b \rar s U$
transition, our next problem is computation of the matrix element of
(\ref{e8701}), (\ref{e8707}) and (\ref{e8708}), between initial and final
state baryons. It follows from Eqs. (\ref{e8701}) and (\ref{e8707}) that the
we need to know the following matrix elements 
\bea
\label{e8712}
&&\lla \Lambda \vel \bar{s} \gamma_\mu b \ver \Lambda_b \rra~, \nnb \\
&&\lla \Lambda \vel \bar{s} \gamma_\mu \gamma_5 b \ver \Lambda_b \rra~.
\eea
These matrix elements can be parametrized in terms of the form factors
a follows \cite{R8711}:
\bea
\label{e8713}
\lla \Lambda \vel \bar{s} \gamma_\mu b \ver \Lambda_b \rra \es
\bar{u}_\Lambda \Big[f_1 \gamma_\mu + f_2 i \sigma_{\mu\nu} q^\nu +
f_3 q_\mu \Big] u_{\Lambda_b}~,\\
\label{e8714}
\lla \Lambda \vel \bar{s} \gamma_\mu \gamma_5 b \ver \Lambda_b \rra \es
\bar{u}_\Lambda \Big[g_1 \gamma_\mu \gamma_5 + g_2 i \sigma_{\mu\nu} q^\nu
\gamma_5 + g_3 q_\mu \gamma_5\Big] u_{\Lambda_b}~, 
\eea
where $q = p_{\Lambda_b} - p_\Lambda$.

It follows from these expressions that $\Lambda_b \rar
\Lambda+missing~energy$ decay is described in terms of numerous form
factors. It is shown in \cite{R8705} that Heavy Effective Quark Theory
(HEQT) reduces the number of independent form factors to two $(F_1$ and
$F_2)$ irrelevant of the Dirac structure of corresponding operators, i.e.,
\bea
\label{e8715}
\lla \Lambda(p_\Lambda) \vel \bar{s} \Gamma b \ver \Lambda_b \rra = 
\bar{u}_\Lambda \Big[F_1(q^2) + \rlap/v F_2(q^2) \Big] \Gamma u_{\Lambda_b}~,
\eea
where $\Gamma$ is the arbitrary Dirac structure and
$v^\mu=p_{\Lambda_b}^\mu/m_{\Lambda_b}$ is the four velocity of $\Lambda_b$.
Comparing Eqs. (\ref{e8713}), (\ref{e8714}) and (\ref{e8715}) one can easily
obtain relations among the form factors (see also \cite{R8711})
\bea
\label{e8716}
f_1 \es g_1 = F_1 + {m_\Lambda \over m_{\Lambda_b}} F_2~, \nnb \\
g_2 \es f_2 = g_3 = f_3 = {F_2\over m_{\Lambda_b}}~,
\eea
which we will use in our numerical analysis.

Using Eqs. (\ref{e8706}), (\ref{e8707}), (\ref{e8708}), (\ref{e8713}),
(\ref{e8714}) and (\ref{e8716}) we get the following expression for the
matrix elements of the $\Lambda_b \rar \Lambda+missing~energy$ decay:
\bea
\label{e8717}
{\cal M}^{(1)} \es {G_F \alpha \over 2 \sqrt{2} \pi} V_{tb} V_{ts}^\ast C_{10}
\, \bar{u}_\Lambda \Big[ \gamma_\mu (f_1-g_1 \gamma_5) +
\rlap/q \gamma_\mu (f_2-g_2 \gamma_5) \Big] u_{\Lambda_b} \,
\bar{\nu} \gamma_\mu (1-\gamma_5) \nu~,\\ \nnb \\
\label{e8718}
{\cal M}^{(2)} \es {1\over \Lambda^{d_{\cal U}}} \, \bar{u}_\Lambda \Big[
A + B \gamma_5 \Big] u_{\Lambda_b} \, O~,\\ \nnb \\ 
\label{e8719}
{\cal M}^{(3)} \es {1\over \Lambda^{d_{\cal U}-1}} \, \bar{u}_\Lambda \Big[
\gamma_\mu (A^V + B^V \gamma_5) + (C^V + D^V \gamma_5) p_{\Lambda_b\mu} 
\Big] u_{\Lambda_b} \, O^\mu~.
\eea  
Here, $i=1$, $i=2$ and $i=3$ correspond to the SM, scalar operator and   
vector operator contributions, respectively, and,
\bea
\label{e8720}
A \es C_S \Big[   (m_{\Lambda_b} - m_\Lambda) f_1 + q^2 f_3 \Big]~,\nnb \\
B \es C_P \Big[ - (m_{\Lambda_b} + m_\Lambda) f_1 + q^2 g_3 \Big]~,\nnb \\
A^V \es C_V \Big[   f_1 - (m_{\Lambda_b} + m_\Lambda) f_2 \Big]~,\nnb \\
B^V \es C_A \Big[   g_1 + (m_{\Lambda_b} + m_\Lambda) g_2 \Big]~,\nnb \\
C^V \es 2 C_V f_2~,\nnb \\
D^V \es 2 C_A g_2~.
\eea

In the derivation of Eqs. (\ref{e8717}), (\ref{e8718}) and (\ref{e8719}), 
the neutrinos are taken to be
massless and we also use $f_3=f_2$ and $g_3=g_2$ ( see Eq. (\ref{e8716}) ). 

Having obtained the matrix elements for the $\Lambda_b \rar
\Lambda+missing~energy$, the differential decay width can be calculated
straightforwardly. As has already been noted, the polarization effects for
the $\Lambda_b \rar \Lambda+missing~energy$, are richer than compared to
the corresponding mesonic decays, since polarization of $\Lambda_b$ and
$\Lambda$ can be measured. In this connection few words about the
polarizations of baryons are in order. In the $\Lambda_b$ rest frame, the
unit vectors along the longitudinal, normal and transversal components of
$\Lambda$ polarization are defined in the following
way:
\bea
\vec{e}_L \es {\vec{p}_\Lambda \over \vel \vec{p}_\Lambda \ver}~, \nnb \\
\vec{e}_N \es \vec{\xi}_{\Lambda_b} \times \vec{e}_L~, \nnb \\
\vec{e}_T \es \vec{e}_L \times \vec{e}_N~,\nnb
\eea
where $\vec{p}_\Lambda$ is the momentum of $\Lambda$ baryon and
$\vec{\xi}_{\Lambda_b}$ is the unit vector along the ${\Lambda_b}$ baryon
spin in its rest frame. In the rest frame of ${\Lambda_b}$ baryon the
differential decay width can be written as:
\bea
\label{e8721}
{d\Gamma^{(i)} \over dE_\Lambda} = \Bigg( {d\Gamma_0^{(i)} \over dE_\Lambda}\Bigg)
{1 \over 4} \Bigg[ 1+ {I_2^{(i)} \over I_1^{(i)} } \vec{e}_L \cdot
\vec{\xi}_{\Lambda_b} \Bigg] \Big[ 1 + \vec{\cal P}_\Lambda^{\;(i)} \cdot
\vec{\xi}_\Lambda \Big] ~,
\eea
where $i=1,2,3$, and $d\Gamma_0^{(i)}/dE_\Lambda$
describes the unpolarized differential decay width.

In Eq. (\ref{e8721})
$\vec{\cal P}_\Lambda^{\;(i)}$ is determined as follows:
\bea
\label{e8722}
\vec{\cal P}_\Lambda^{\;(i)} = {1 \over 1 + {\ds {I_2^{(i)} \over I_1^{(i)} } }
\vec{e}_L \cdot \vec{\xi}_{\Lambda_b} } \Bigg[ \Bigg( 
{I_3^{(i)} \over I_1^{(i)} } + {I_4^{(i)} \over I_1^{(i)} } \vec{e}_L \cdot
\vec{\xi}_{\Lambda_b} \Bigg) \vec{e}_L
+ {I_5^{(i)} \over I_1^{(i)} } \vec{e}_T + {I_6^{(i)} \over I_1^{(i)} }
\vec{e}_N \Bigg]~.
\eea

Note that $\Lambda_b \rar \Lambda \bar{\nu} \nu$ decay, with
$\Lambda_b$ and $\Lambda$ polarizations is studied in \cite{R8711}. Explicit
expressions of $d\Gamma_0/dE_\Lambda$, $\vec{\cal P}_\Lambda$, 
$I_2$ and $I_1$ in the SM are given in \cite{R8711}, and therefore we do not 
present them in this work.

After simple calculation, the decay width due to the scalar operator takes
the following form:
\bea
\label{e8723} 
{d\Gamma_0^{(2)} \over dE_\Lambda} = {1 \over 2 m_{\Lambda_b}} {A_{d_{\cal U}} \over \ga
\Lambda_{\cal U}^{d_{\cal U}}\dr^2 } \ga q^2 \dr^{d_{\cal U}-2} {\vel \vec{p}_\Lambda \ver \over
(2\pi)^2 } I_1^{(2)}~,
\eea
where $q=p_{\Lambda_b}-p_\Lambda$, and $\vel \vec{p}_\Lambda \ver$ is the 
magnitude of the $\Lambda_b$ baryon three momentum, and  
\bea
\label{e8724}
I_1^{(2)} \es 4 m_{\Lambda_b} \Big[ \vel A \ver^2 \ga E_\Lambda + m_\Lambda \dr + \vel
B \ver^2 \ga E_\Lambda - m_\Lambda \dr \Big]~, \nnb \\
I_2^{(2)} \es - 8 {\rm Re}[A B^\ast ] m_{\Lambda_b} 
\vel \vec{p}_\Lambda \ver~, \nnb \\
I_3^{(2)} \es - 8 {\rm Re}[A B^\ast ] m_{\Lambda_b} 
\vel \vec{p}_\Lambda \ver \equiv I_2^{(2)}~, \nnb \\
I_4^{(2)} \es 4 \vel A \ver^2 \Big[ - \vel \vec{p}_\Lambda \ver^2 +
m_\Lambda \ga E_\Lambda + m_\Lambda \dr \Big] m_{\Lambda_b} +
4 \vel B \ver^2 \Big[ \vel \vec{p}_\Lambda \ver^2 -
m_\Lambda \ga E_\Lambda - m_\Lambda \dr \Big] m_{\Lambda_b}~, \nnb \\
I_5^{(2)} \es 4 \vel A \ver^2 m_{\Lambda_b} \ga E_\Lambda + m_\Lambda \dr - 
4 \vel B \ver^2 m_{\Lambda_b} \ga E_\Lambda - m_\Lambda \dr~, \nnb \\
I_6^{(2)} \es 8 \vel \vec{p}_\Lambda \ver m_{\Lambda_b} 
{\rm Im} [A B^\ast ]~.
\eea
The coefficients $\vec{e}_L$, $\vec{e}_T$ and $\vec{e}_N$, in Eq.
(\ref{e8722}) corresponding
to the longitudinal, transversal and normal polarization asymmetries of
$\Lambda$ can also be defined as:
\bea
{\cal P}_{\Lambda,j}^{(i)} = {\ds d\Gamma^{(i)}(\vec{\xi}_\Lambda\cdot \vec{e}_j=1) - 
                       d\Gamma^{(i)}(\vec{\xi}_\Lambda\cdot \vec{e}_j=-1)
\over \ds
           d\Gamma^{(i)}(\vec{\xi}_\Lambda\cdot \vec{e}_j=1) + 
           d\Gamma^{(i)}(\vec{\xi}_\Lambda\cdot \vec{e}_j=-1)}~,\nnb
\eea
where $j=L,T,N$. 

Performing similar calculations for unpolarized decay due to the vector
operator, we get:
\bea
{d\Gamma_0^{(3)} \over d E_\Lambda} = {1\over 2 m_{\Lambda_b}} {A_{d_{\cal U}} \over
\ga \Lambda_{\cal U}^{d_{\cal U}-1} \dr^2 }  \ga q^2\dr^{d_{\cal U}-2} {\vel \vec{p}_\Lambda \ver
\over (2\pi)^2 } I_1^{(3)}~,\nnb
\eea
and the expressions of the functions entering into Eq. (\ref{e8722}) 
for the vector operator case are as follows:
\bea
I_1^{(3)} \es 4 {\vel D_V \ver^2 \over q^2} m_{\Lambda_b}^3 (E_\Lambda -
m_\Lambda) \Big[ (E_\Lambda - m_{\Lambda_b})^2 - q^2 \Big] \nnb \\
\ar 4 {\vel C_V \ver^2 \over q^2} m_{\Lambda_b}^3 (E_\Lambda +
m_\Lambda) \Big[ (E_\Lambda - m_{\Lambda_b})^2 - q^2 \Big] \nnb \\
\ar 4 {\vel A_V \ver^2 \over q^2} m_{\Lambda_b} \Big[ -2 m_{\Lambda_b} 
E_\Lambda^2 + E_\Lambda (2 m_\Lambda^2 + 2 m_{\Lambda_b}^2 + q^2) - 
m_\Lambda (2 m_\Lambda m_{\Lambda_b} + 3 q^2) \Big] \nnb \\
\ar 4 {\vel B_V \ver^2 \over q^2} m_{\Lambda_b} \Big[ -2 m_{\Lambda_b} 
E_\Lambda^2 + E_\Lambda (2 m_\Lambda^2 + 2 m_{\Lambda_b}^2 + q^2) - 
m_\Lambda (2 m_\Lambda m_{\Lambda_b} - 3 q^2) \Big] \nnb \\
\ar 8 {\mbox{\rm Re}[A_V C_V^\ast ]\over q^2} m_{\Lambda_b}^2 (E_\Lambda +
m_\Lambda) \Big[ (E_\Lambda - m_{\Lambda_b}) (m_\Lambda - m_{\Lambda_b})
-q^2 \Big] \nnb \\
\ar 8 {\mbox{\rm Re}[B_V D_V^\ast]\over q^2} m_{\Lambda_b}^2 (E_\Lambda - 
m_\Lambda) \Big[ (E_\Lambda - m_{\Lambda_b}) (m_\Lambda + m_{\Lambda_b})
+q^2 \Big] \nnb \\ \nnb \\
I_2^{(3)} \es - 8 {\vel \vec{p}_\Lambda \ver \over q^2} \Big\{
\mbox{\rm Re}[C_V D_V^\ast] m_{\Lambda_b}^3 \Big[(E_\Lambda - m_{\Lambda_b})^2 -
q^2 \Big] \nnb \\
\ar \mbox{\rm Re}[A_V B_V^\ast] m_{\Lambda_b} \Big[ 2 m_\Lambda^2 -
2 m_{\Lambda_b} E_\Lambda + q^2 \Big] +
\mbox{\rm Re}[B_V C_V^\ast] m_{\Lambda_b}^2 \Big[ (E_\Lambda -
m_{\Lambda_b}) (m_\Lambda + m_{\Lambda_b}) \Big] \nnb \\
\ar \mbox{\rm Re}[A_V D_V^\ast] m_{\Lambda_b}^2 \Big[ (E_\Lambda -
m_{\Lambda_b}) (m_\Lambda + m_{\Lambda_b}) - q^2\Big]\Big\}~, \nnb \\ \nnb \\
I_3^{(3)} \es {8\over q^2} \Big\{- \mbox{\rm Re}[C_V D_V^\ast] m_{\Lambda_b}^3
\Big[(E_\Lambda - m_{\Lambda_b})^2 - q^2 \Big] \nnb \\
\ek \mbox{\rm Re}[A_V D_V^\ast] m_{\Lambda_b}^2 \Big[ (E_\Lambda - 
m_{\Lambda_b}) (m_\Lambda - m_{\Lambda_b}) - q^2 \Big] \nnb \\
\ek \mbox{\rm Re}[B_V C_V^\ast] m_{\Lambda_b}^2 \Big[ (E_\Lambda -
m_{\Lambda_b}) (m_\Lambda + m_{\Lambda_b}) - q^2\Big] \nnb \\
\ar \mbox{\rm Re}[A_V B_V^\ast] m_{\Lambda_b} \Big[ 2 m_{\Lambda_b}
(m_{\Lambda_b} - E_\Lambda) + q^2 \Big] \Big\}~, \nnb \\ \nnb \\
I_4^{(3)} \es {4\over q^2 m_\Lambda} \Big\{ \vel D_V \ver^2 m_{\Lambda_b}^3 
\Big[ \vel \vec{p}_\Lambda \ver^2 + m_\Lambda (m_\Lambda - E_\Lambda) \Big]
\Big[(E_\Lambda - m_{\Lambda_b})^2 - q^2)\Big] \nnb \\
\ar \vel C_V \ver^2 m_{\Lambda_b}^3 
\Big[ - \vel \vec{p}_\Lambda \ver^2 + m_\Lambda (m_\Lambda + E_\Lambda) \Big]
\Big[(E_\Lambda - m_{\Lambda_b})^2 - q^2)\Big] \nnb \\
\ar 2 \mbox{\rm Re}[A_V C_V^\ast ] m_{\Lambda_b}^2 \Big[ - \vel
\vec{p}_\Lambda \ver^2 + m_\Lambda (m_\Lambda + E_\Lambda) \Big]
\Big[(E_\Lambda - m_{\Lambda_b}) (m_\Lambda - m_{\Lambda_b}) - q^2\Big] \nnb \\
\ar 2 \mbox{\rm Re}[B_V D_V^\ast ] m_{\Lambda_b}^2 \Big[ \vel
\vec{p}_\Lambda \ver^2 + m_\Lambda (m_\Lambda - E_\Lambda) \Big]
\Big[(E_\Lambda - m_{\Lambda_b}) (m_\Lambda + m_{\Lambda_b}) + 
q^2\Big] \nnb \\
\ar \vel B_V \ver^2 m_{\Lambda_b} \Big[ 2 m_\Lambda (E_\Lambda -
m_{\Lambda_b}) (m_{\Lambda_b} E_\Lambda - m_\Lambda ^2) + 
2 \vel \vec{p}_\Lambda \ver^2 \Big( m_\Lambda^2 + m_\Lambda m_{\Lambda_b} +
m_{\Lambda_b} (m_{\Lambda_b} - E_\Lambda) \Big) \nnb \\
\ek q^2 \Big( \vel \vec{p}_\Lambda \ver^2 + m_\Lambda (m_\Lambda -
E_\Lambda) \Big) \Big] \nnb \\
\ek \vel A_V \ver^2 m_{\Lambda_b} \Big[ \vel \vec{p}_\Lambda \ver^2 \Big(
2 m_\Lambda^2 - 2 m_{\Lambda_b} (E_\Lambda + m_\Lambda) + 2 m_{\Lambda_b}^2
- q^2 \Big) \nnb \\
\ar m_\Lambda \Big( 2 E_\Lambda^2 m_{\Lambda_b} + 2 m_\Lambda^2 m_{\Lambda_b} +
m_\Lambda q^2 - 2 E_\Lambda (m_\Lambda^2 + m_{\Lambda_b}^2) +  E_\Lambda q^2
\Big) \Big] \Big\}~, \nnb \\ \nnb \\
I_5^{(3)} \es - 4 \vel B_V \ver^2 m_{\Lambda_b} (E_\Lambda - m_\Lambda) +
4 \vel A_V \ver^2 m_{\Lambda_b} (E_\Lambda + m_\Lambda)~.\nnb
\eea

For the case when $\Lambda_b$ is unpolarized, we get from Eq. (\ref{e8722}) that
\bea
\vec{\cal P}_\Lambda^{\;(i)} = \alpha_\Lambda^{(i)} \vec{e}_L~, \nnb
\eea
with
\bea
\alpha_\Lambda^{(i)} = {I_3^{(i)} \over I_1^{(i)}}~,\nnb
\eea
which means that, in this case $\Lambda$ polarization is purely
longitudinal.

For $\Lambda$ unpolarized, by performing summation
over $\Lambda$ spin in Eq. (\ref{e8722}), we get
\bea
{d\Gamma^{(i)} \over dE_\Lambda} =  \Bigg({d\Gamma_0^{(i)} \over
dE_\Lambda} \Bigg) {1\over 2}\Big[ 1 + \alpha_{\Lambda_b}^{(i)} 
\vec{\xi}_{\Lambda_b} \cdot \vec{e}_L \Big]~,\nnb
\eea
where 
\bea
\alpha_{\Lambda_b}^{(i)} = {I_2^{(i)} \over I_1^{(i)} }~.\nnb
\eea

Note that the normal component ${\cal P}_N^{(i)}$ of $\Lambda$ polarization is a
T--odd quantity and its non--zero value indicates CP violation. In the SM and
considered version of unparticle physics, there is no CP violating phase (in
the SM case $V_{tb} V_ts^\ast$), therefore, it cannot induce ${\cal P}_N$ in    
$\Lambda_b \rar \Lambda+missing~ energy$ decay when both baryons are
polarized. Obviously, if ${\cal P}_N$ is measured in experiments, it clearly is an
indication of the fact that there do not exist new CP violating sources.

\section{Numerical analysis}

In this section we calculate the numerical values of the differential
branching ratio and polarizations of $\Lambda_b \rar \Lambda \bar{\nu} \nu$
decay in unparticle physics.

The transition form factors $f_i$ and $g_i$, as well as $d_{\cal U}$ and
$\Lambda_{\cal U}^{d_{\cal U}}$, are the main input parameters in the numerical analysis. For
calculation of the form factors we use the results obtained from QCD sum
rules method in corporation with HQET. As has already been noted, HQET
reduces the number of independent form factors to two. Moreover, the $q^2$
dependence of $F_i$ which is obtained in terms of three--parameter fit has
the form \cite{R8711}
\bea
F_i(q^2) = {\ds F_i(0) \over \ds 1 - a_F^i (q^2/m_{\Lambda_b}^2) + b_F^i
(q^2/m_{\Lambda_b}^2)^2}~, \nnb
\eea
The values of $F_i(0)$, $a_F^i$ and $b_F^i$ are given in Table--1.

\begin{table}[h]
\renewcommand{\arraystretch}{1.5}
\addtolength{\arraycolsep}{3pt}
$$
\begin{array}{|l|c|c|c|}
\hline
& F(0) & a_F & b_F \\ \hline
F_1(0) &
\phantom{-}0.462 & -0.0182 & -0.000176 \\ \hline
F_2(0) &
-0.077 & -0.0685 & \phantom{-}0.001460 \\ \hline
\end{array}   
$$
\caption{Form factors for $\Lambda_b \rar \Lambda+missing~ energy$ decay in
a three parameter fit.}
\renewcommand{\arraystretch}{1}
\addtolength{\arraycolsep}{-3pt}
\end{table}   

It is emphasized in \cite{R8714} that unparticles behave as a non--integer
number of particles and it is shown there that the very peculiar shape of
u--quark energy distribution in the $t \rar c {\cal U}$ decay and it 
can serve as a good test in discovering
unparticles experimentally. Along the same lines, the energy distribution of
$K$ and $K^\ast$ mesons in the $B \rar K(K^\ast)+missing~ energy$ decay is
analyzed \cite{R8733} and it is seen that this decay, especially in the
presence of vector unparticle operators, is very distinctive compared to that 
of the SM prediction. Similar situation can take place for the $\Lambda_b
\rar \Lambda+missing~ energy$ decay. In what follows, we try to answer the
intriguing question whether the polarization observables can be useful 
for the experimental observation of unparticles.

It is shown in \cite{R8731} that if vector operators couple to the flavor
non--diagonal current, $d_{\cal U}$ should be larger than 
$d_{\cal U}>2$. On the other hand, the bound for the scalar operators
turns out to be $d_{\cal U}>1$, and these are the bounds we will use in our numerical
analysis. The values of other parameters are chosen as
$C_P=C_S=2.0\times 10^{-3}$ for scalar operators; and $C_V=C_A=10^{-5}$ 
for vector operator, and $\Lambda_{\cal U}=1~TeV$ for both cases.

In Figs. (1) and (2), we present the dependence of the differential decay
width as a function of the $\Lambda$ baryon energy $E_\Lambda$ for scalar
and vector operators, for various choices of $d_{\cal U}$, respectively. From these
figures we see that the distribution for the final $\Lambda$ baryon energy
for both operators are similar. Note that, the behavior of the dependence 
of the differential decay width on the energy distribution $E_\Lambda$ for
these operators and SM case are very similar to each other.

In Figs. (3) and (4) we present the dependence of the branching ratios of
the $\Lambda_b \rar \Lambda+missing~ energy$ decay on $d_{\cal U}$ at fixed values
of the effective coupling constants $C_S$, $C_P$, $C_V$ and $C_A$, at
different values of the cut--off scale $\Lambda_{\cal U}$, for scalar and vector
operators. For completeness, in these figures we also present the SM result
for the branching ratio of the $\Lambda_b \rar \Lambda \bar{\nu} \nu$ decay.
From these figures we see
that, for $d_{\cal U}>2$, the value of the branching ratio is smaller compared to
that of the SM case in the presence of the vector operator; while the
branching ratio can exceed the SM prediction in the presence of the scalar
operator when $d_{\cal U}<1.7$, whose behavior is determined by 
$\Lambda_{\cal U}$. Therefore,
determination of the value of the branching ratio can put stringent 
restrictions to the values of $d_{\cal U}$ and $\Lambda_{\cal U}$.

In Figs. (5) and (6), we present the dependence of the branching ratio of
the $\Lambda_b \rar \Lambda+missing~ energy$ decay on cut--off scale
$\Lambda_{\cal U}$ at fixed values of 
$C_S$, $C_P$ $C_V$ and $C_A$, respectively. We observe from these figures that, 
the branching ratios are very sensitive to the values of these effective
couplings. It follows from Fig. (5) that, for the scalar operator case and
up to $\Lambda_{\cal U} = 8~TeV$, the value of the branching ratio exceeds that 
of the SM prediction, for all choices of the fixed values of $d_{\cal U}$,
which can give useful information about unparticle physics.

Now let us discuss the numerical values of the $\Lambda$ and $\Lambda_b$
baryon polarizations. As we proceed in analyzing the $\Lambda_b$
polarizations we have assumed that $\Lambda$ is not polarized, and when we
analyze $\Lambda$ polarizations we have assumed that
$\Big(I_2^{(i)}/I_1^{(i)}\Big) \vec{e}_L\cdot \vec{\xi}_{\Lambda_b}$ is
small which can be neglected in numerical calculations.  

The results we get for $\alpha_{\Lambda_b} = I_2/I_1$ can be summarized as
follows:

\begin{itemize}

\item In the presence of the scalar operator the magnitude of
$\alpha_{\Lambda_b}$ starts from zero and approaches sharply to the value 
one; and from $E_\Lambda \simeq 1.25~GeV$ on, its value continues to be very 
close to one. It is further observed that in the whole physical region of
$E_\Lambda$, its value is independent of the values of the parameters $C_P$
and $C_S$. Averaged value of $\alpha_{\Lambda_b}$ in unparticle physics
model is equal to $0.98$.

\item In the vector operator case the situation is drastically different
from that above--mentioned scalar operator case. In other words,
in the region $\ga E_\Lambda\dr_{min} \le E_\Lambda \le 1.70~GeV$,
$\alpha_{\Lambda_b}$ gets negative values and at $E_\Lambda = 1.70~GeV$, 
$\alpha_{\Lambda_b}$ becomes zero. Starting from $E_\Lambda = 1.70~GeV$ on, 
$\alpha_{\Lambda_b}$ increases with the increasing values of $E_\Lambda$. 
Similar situation occurs
for the SM case as well. (see \cite{R8711}). More essential than that, similar to
the scalar operator case, $\alpha_{\Lambda_b}$ is insensitive to the values
of the parameters $C_P$ and $C_A$. Note that  in the SM 
$\lla \alpha_{\Lambda_b}\rra=-0.33$, while this model predicts 
$\lla \alpha_{\Lambda_b}=0.86\rra$.
\end{itemize}

From the analysis of $\Lambda$ polarizations we get the following results:

\begin{itemize}

\item In the presence of scalar and vector operators, longitudinal polarization 
of $\Lambda$ exhibits practically the same behavior, namely, up to $E_\Lambda
= 1.25~GeV$, ${\cal P}_L$ increases and from that point on it remains constant for
all kinematical region, and its value is independent of the parameters
$C_S$, $C_P$,$C_V$ and $C_A$. 

\item The averaged value of the longitudinal polarization is $\lla {\cal P}_L \rra
= 0.98$ in the scalar and $\lla {\cal P}_L \rra  = 0.99$ in the vector operator
cases, respectively. Note that $\lla {\cal P}_L \rra  \approx -0.33$ in the SM
case.

\item Transversal polarization of $\Lambda$ decreases with increasing values 
of $E_\Lambda$ and from $E_\Lambda=1.75~GeV$ on it approaches to zero in the
presence of the scalar operator, being insensitive to the parameters $C_S$
and $C_P$. This behavior is very different in the vector operator case. While 
the sign of ${\cal P}_T$ is negative up to $E_\Lambda = 1.25~GeV$, it starts
increasing from this point on, and attains at constant value $10\%$ after
$E_\Lambda = 1.5~GeV$. Similar to the scalar operator case, ${\cal P}_T$ is
insensitive to the numerical values of $C_V$ and $C_A$.

\item The averaged value of the transversal polarization is $\lla {\cal P}_T \rra =
7\%$ in the vector and $\lla {\cal P}_T \rra = ~4.6\%$ in the scalar operator
case, respectively, and $\lla {\cal P}_T \rra = ~5.4\%$ in the SM case.

In conclusion, we have studied the possible manifestation of unparticles 
on the missing energy signatures of the rare $\Lambda_b$ decays. The
branching ratio, $\Lambda$ and $\Lambda_b$ baryon polarizations are studied
in unparticle physics in the presence of scalar and vector operators. The
value of the branching ratio is very sensitive to the choice of the model
parameter $d_{\cal U}$. The averaged values of $\Lambda$ and $\Lambda_b$
baryon polarizations significantly differ from the corresponding SM values.
Therefore experimental measurements of the branching ratio of the process
$\Lambda_b \rar \Lambda+missing~ energy$ decay, as well as $\Lambda$ and
$\Lambda_b$ baryon polarizations can give useful information on unparticle
physics.

\end{itemize}

\newpage

\section*{Figure captions}
{\bf Fig. (1)} The dependence of the unpolarized differential decay width 
on $\Lambda$ baryon energy at $\Lambda_{\cal U}= 1~TeV$, and at fixed values
of $d_{\cal U}$ for the scalar operator. \\ \\
{\bf Fig. (2)} The same as in Fig. (1), but for the vector operator. \\ \\
{\bf Fig. (3)} The dependence of the branching ratio of $\Lambda_b \rar
\Lambda+missing~ energy$ decay on $d_{\cal U}$ at fixed values of
$\Lambda_{\cal U}$ for the scalar operator. \\ \\
{\bf Fig. (4)} The same as in Fig. (3), but for the vector operator. \\ \\
{\bf Fig. (5)} The dependence of the branching ratio of $\Lambda_b \rar
\Lambda+missing~ energy$ decay on $\Lambda_{\cal U}$ at fixed values of
$d_{\cal U}$ for the scalar operator. \\ \\
{\bf Fig. (6)} The same as in Fig. (5), but for the vector operator.

\newpage

\begin{figure}  
\vskip 4.0 cm   
    \includegraphics{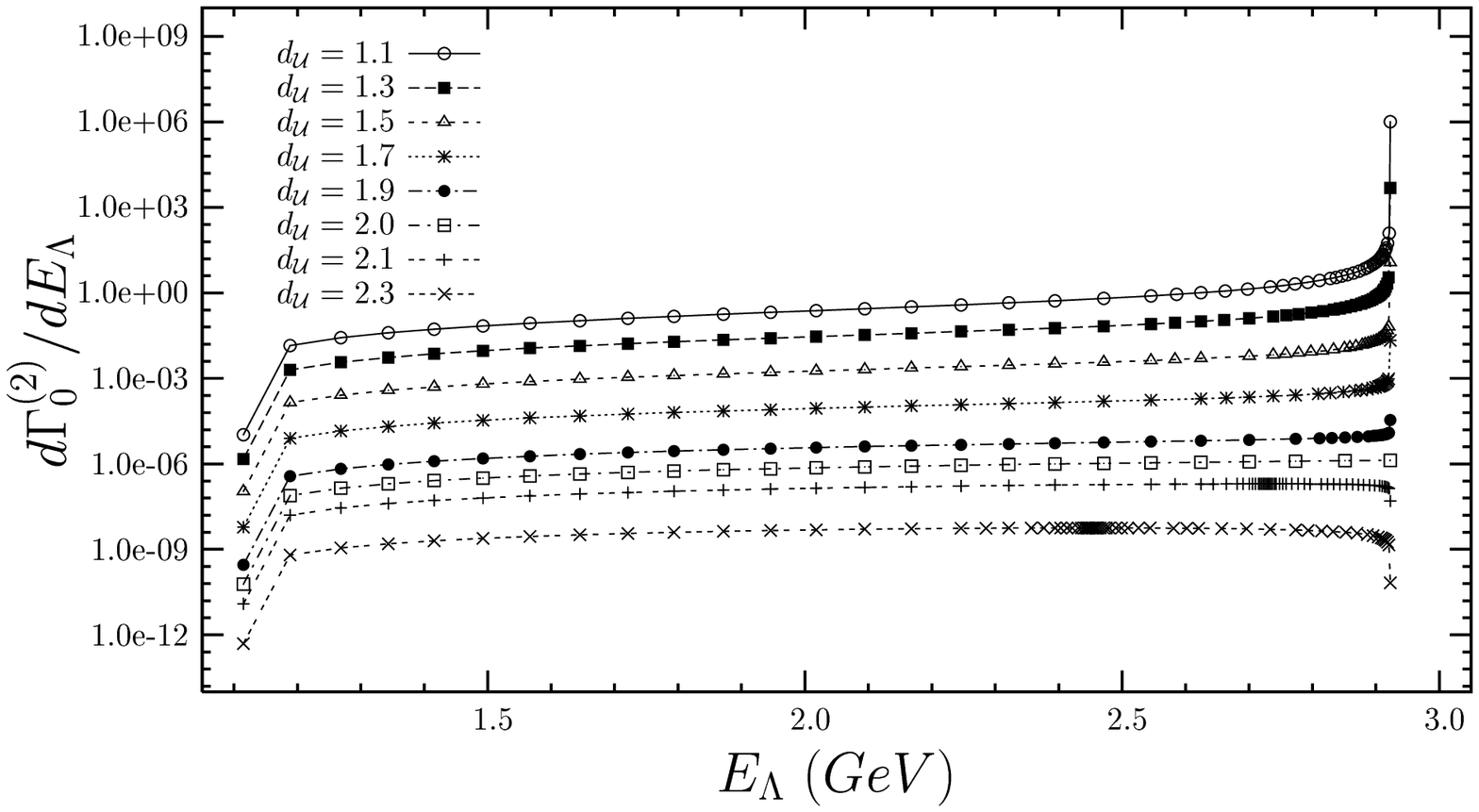}
\vskip 6.3 cm
\caption{}
\end{figure}

\begin{figure}  
\vskip 4.0 cm
    \includegraphics{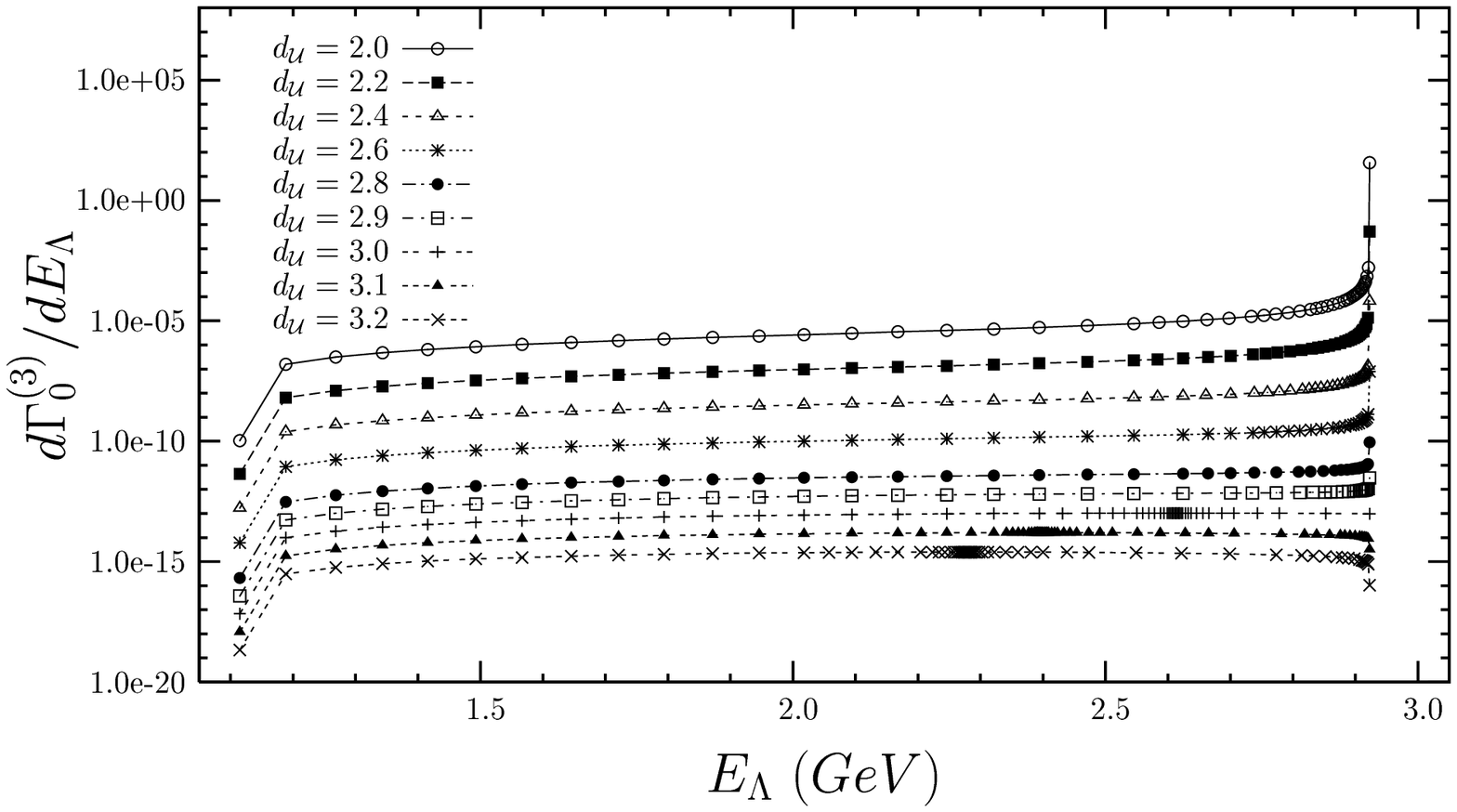}
\vskip 6.3 cm
\caption{}
\end{figure}

\begin{figure}
\vskip 4.0 cm
    \includegraphics{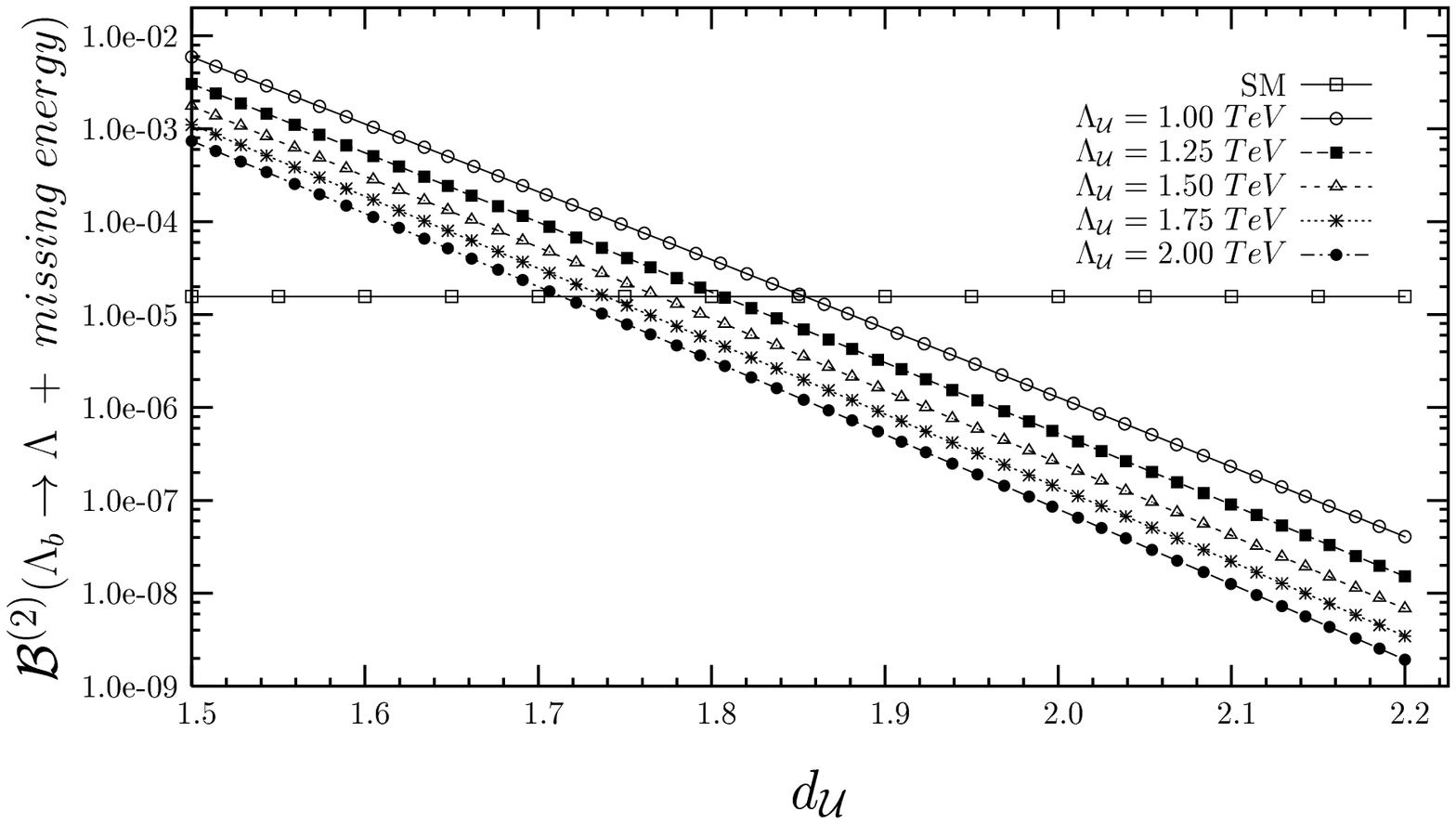}
\vskip 6.3 cm
\caption{}      
\end{figure}

\begin{figure}
\vskip 4.0 cm
    \includegraphics{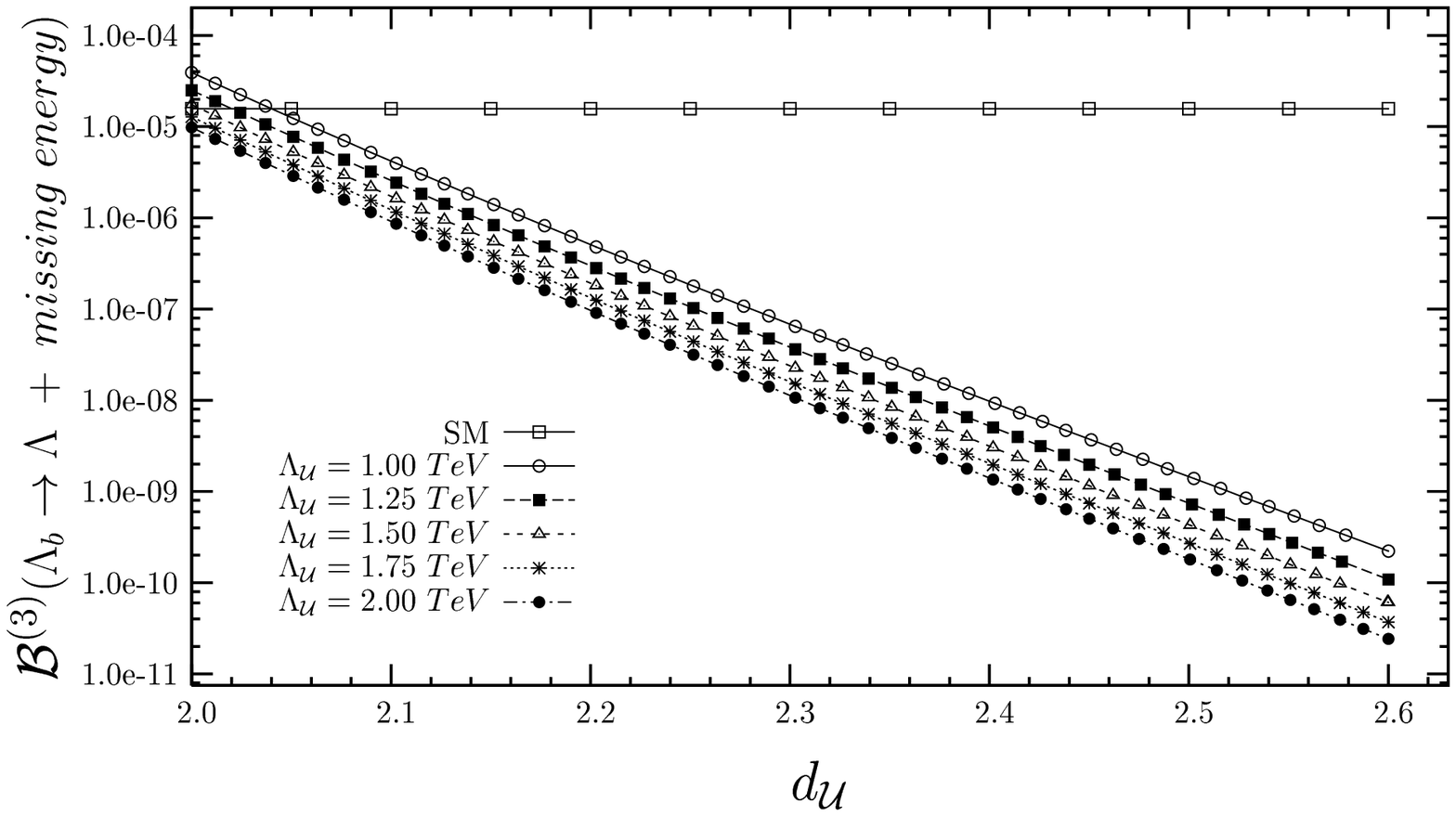}
\vskip 6.3 cm
\caption{}      
\end{figure}

\begin{figure}
\vskip 4.0 cm
    \includegraphics{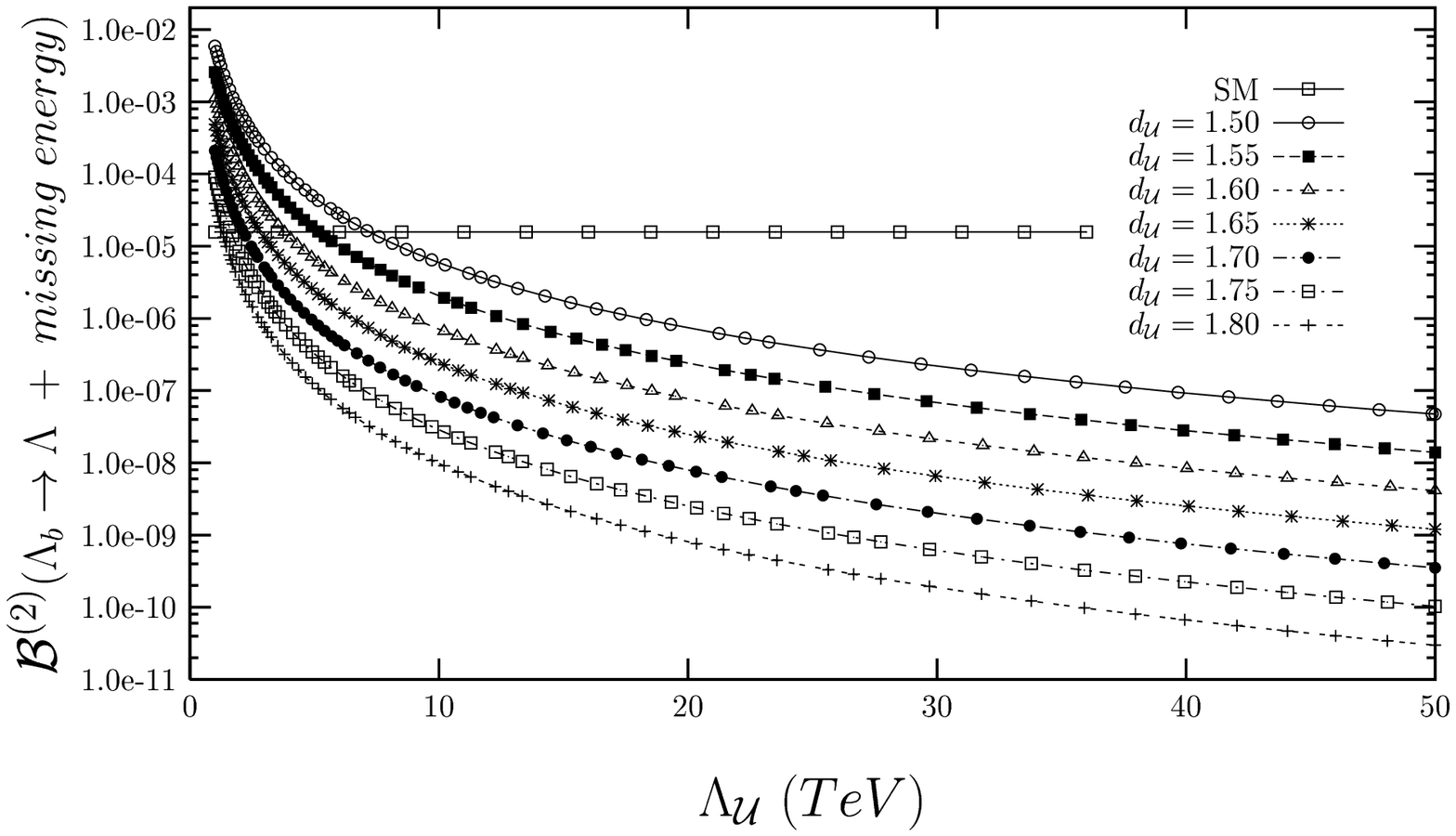}
\vskip 6.3 cm
\caption{}      
\end{figure}

\begin{figure}
\vskip 4.0 cm
    \includegraphics{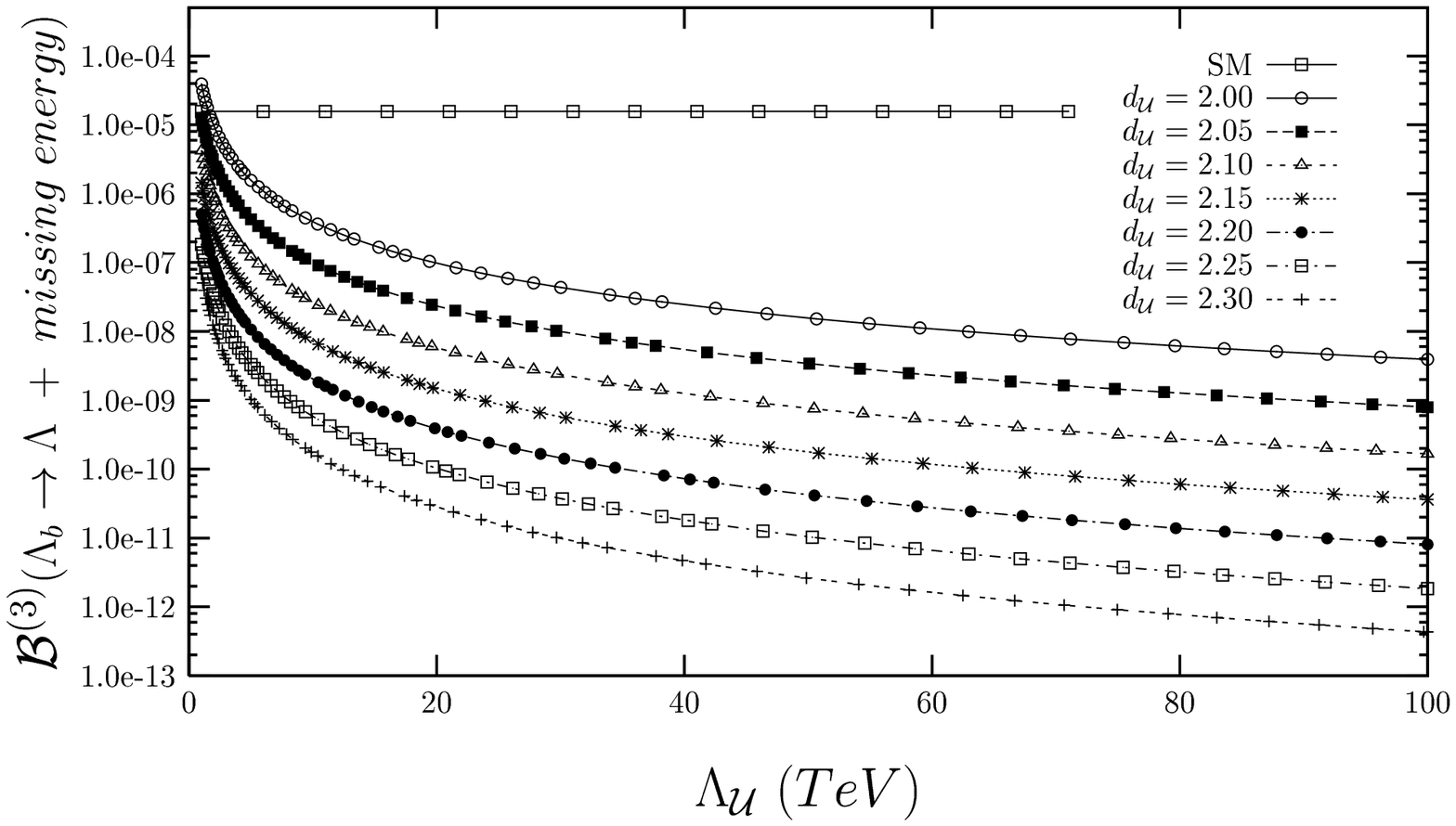}
\vskip 6.3 cm
\caption{}      
\end{figure}






\end{document}